\documentclass[%
 ,superscriptaddress%
,amssymb, amsmath,nobibnotes, aip, pop]{revtex4-1}

\usepackage{graphicx}

\begin{document}

\title{Chaotic saddles in nonlinear modulational interactions in a plasma}%

\author{Rodrigo A. Miranda}
\email{rmiracer@unb.br}
\affiliation{Institute of Aeronautical Technology (ITA) and World Institute for Space Environment Research (WISER), S\~ao Jos\'e dos Campos-SP, 12228-900 Brazil.}
\affiliation{National Institute for Space Research (INPE) and World Institute for Space Environment Research (WISER), P. O. Box 515, S\~ao Jos\'e dos Campos-SP, 12227-010, Brazil.}
\affiliation{University of Bras\'ilia (UnB), Gama Campus, and Plasma Physics Laboratory, Institute of Physics, Bras\'ilia-DF 70910-900, Brazil.}
\author{Erico L. Rempel}
\affiliation{Institute of Aeronautical Technology (ITA) and World Institute for Space Environment Research (WISER), S\~ao Jos\'e dos Campos-SP, 12228-900 Brazil.}
\affiliation{National Institute for Space Research (INPE) and World Institute for Space Environment Research (WISER), P. O. Box 515, S\~ao Jos\'e dos Campos-SP, 12227-010, Brazil.}
\author{Abraham C.-L. Chian}
\affiliation{Institute of Aeronautical Technology (ITA) and World Institute for Space Environment Research (WISER), S\~ao Jos\'e dos Campos-SP, 12228-900 Brazil.}
\affiliation{National Institute for Space Research (INPE) and World Institute for Space Environment Research (WISER), P. O. Box 515, S\~ao Jos\'e dos Campos-SP, 12227-010, Brazil.}
\affiliation{Observatoire de Paris, LESIA, CNRS, 92195 Meudon, France.}

\begin{abstract}
A nonlinear model of modulational processes in the subsonic regime involving a linearly unstable wave and two linearly damped waves with different damping rates in a plasma is studied numerically. We compute the maximum Lyapunov exponent as a function of the damping rates in a two-parameter space, and identify shrimp-shaped self-similar structures in the parameter space. By varying the damping rate of the low-frequency wave, we construct bifurcation diagrams and focus on a saddle-node bifurcation and an interior crisis associated with a periodic window. We detect chaotic saddles and their stable and unstable manifolds, and demonstrate how the connection between two chaotic saddles via coupling unstable periodic orbits can result in a crisis-induced intermittency. The relevance of this work for the understanding of modulational processes observed in plasmas and fluids is discussed.

\end{abstract}

\maketitle

\section{Introduction}

The theory of nonlinear modulational processes can explain the generation of modulated wave envelopes commonly observed in laboratory plasmas \cite{kim_etal:1974,amiranoff_etal:1992}, space and astrophysical plasmas\cite{ergun_etal:1991,bonnell_etal:1997,weatherall:1997} and oceans \cite{janssen:2003}. For example, modulational Langmuir processes in plasmas refer to the mechanism by which initially small fluctuations of electrostatic Langmuir waves produce density accumulation towards regions where the electric field is weaker. The high-frequency Langmuir waves become trapped in cavitons where the plasma density is lower, resulting in the formation of Langmuir envelope solitons. If the phase velocity of the density perturbation is lower than the ion-acoustic velocity, this process is known as a subsonic modulational process \cite{goldman:1984}. Modulational processes have been observed in laboratory experiments of non-magnetized plasmas in the presence of an external electric field \cite{kim_etal:1974} and in Nd-laser beat-wave experiments in a homogeneous plasma \cite{amiranoff_etal:1992}. In space plasmas, modulational processes can explain the detection of modulated wave envelopes observed in the auroral region \cite{ergun_etal:1991,bonnell_etal:1997}.

Theoretical models of modulational processes in astrophysical plasmas have been derived and studied numerically by several authors. For example, \citet{pataraya_melikidze:1980} used the nonlinear Schr\"odinger (NLS) equation with nonlinear Landau damping to study the modulational instability of nonlinear waves in relativistic plasmas. \citet{weatherall:1997} studied a nonlinear model of plasma waves in pulsars, and found that the onset of wave turbulence occurs via a modulational instability. \citet{robinson_etal:2002} proposed a unifying framework which incorporates modulational and decay processes driven by a pump Langmuir wave. They performed analytical and numerical studies to obtain quantities such as growth rates, stability boundaries and instability thresholds for various types of modulational and decay processes. Recently, \citet{asenjo_etal:2012} derived a NLS equation to describe the self-modulation of nonlinear waves in a hot magnetized relativistic electron-positron plasma.

In the weakly nonlinear regime, modulational processes in plasmas and fluids can be described by low-dimensional models of nonlinear wave-wave interactions. A great amount of literature has been dedicated to their study, for example, \citet{russell_ott:1981} performed a numerical study of a three-wave truncation derived from the NLS. By varying a control parameter, they found a variety of dynamical behaviors such as transitions from stable stationary points to periodic orbits via Hopf bifurcations, period-doubling bifurcations leading to chaos, and transitions from chaotic to periodic solutions via tangent bifurcations. \citet{gosh_papadopoulos:1987} showed that the model of \citet{russell_ott:1981} can also apply to a three-wave truncation of the derivative nonlinear Schr\"odinger equation (DNLS) to study the onset of Alfv\'enic turbulence. \citet{dubois_etal:1985} performed Monte Carlo simulations of a truncated model to study the transition from the sonic to subsonic regime of the modulational instability. \citet{chian_abalde:1997} derived a nonlinear model of the hybrid stimulated modulational instability where Langmuir, electromagnetic and ion-acoustic waves are coupled, and its temporal dynamics was studied by varying several parameters such as the frequency mismatch, dissipation rate and wave dispersion. In \citet{miranda_etal:2005} two different types of intermittency were found in chaotic time series obtained from a three-wave truncation of the NLS. Chaos in a three-wave truncation model of the derivative nonlinear Schr\"odinger equation for nonlinear Alfv\'en waves has been studied by \citet{sanmartin_etal:2004} and \citet{sanchez-arriaga_etal:2007, sanchez-arriaga_etal:2009a, sanchez-arriaga_etal:2009b}. \citet{mori_janssen:2006} derived an expression relating the kurtosis of the sea surface elevation and the occurrence probability of freak waves generated via nonlinear four-wave modulational interactions. Their statistical theory is in good agreement with observational data of extreme wave events \cite{mori_janssen:2006,mori_etal:2007}.

In this paper we study a low-dimensional model of nonlinear four-wave subsonic modulational processes. We start our numerical analysis by exploring a two-dimensional (2D) parameter space in terms of the maximum Lyapunov exponent. A number of papers have shown the existence of regular, self-similar structures in 2D parameter space corresponding to periodic oscillations, embedded within regions corresponding to chaotic dynamics. These structures are called ``shrimps'' due to their characteristic shape consisting of a main central body and four elongated ``antennae'' or ``legs'' \cite{gallas:1993,zou_etal:2010}, and are globally organized along straight lines, ellipses, and spirals. A theoretical explanation of the spiral organization of shrimps was presented by \citet{stoop_etal:2010}. They performed numerical simulations and laboratory experiments of the Nishio-Inaba electronic circuits to demonstrate that the spiral emergence of shrimps is due to the presence of a homoclinic saddle-focus point in the parameter space. Shrimps have been found in several models including discrete-time maps \cite{gallas:1993,lorenz:2008}, continuous-time models represented by ordinary differential equations \cite{bonatto_etal:2003,zou_etal:2010} and in a number of experiments using nonlinear circuits \cite{maranhao_etal:2008,stoop_etal:2010}. Our analysis is then focused on a periodic window, in which we perform a numerical detection of chaotic saddles, which are nonattracting chaotic sets responsible for chaotic transients in nonlinear systems \cite{grebogi_etal:1983,kantz_grassberger:1985,hsu_etal:1988,lai_tel:2011}. They also play an important role in the intermittent dynamics of low- and high-dimensional systems as demonstrated by recent papers \cite{chian_etal:2010,rempel_etal:2004a,rempel_chian:2007}, hence chaotic saddles are of fundamental importance for the characterization of the system dynamics.

This paper is organized as follows. In section \ref{section_model} we review the derivation of the nonlinear model used in this work. In section \ref{section_algorithms} we describe the main numerical algorithms used for the computation of the Lyapunov exponents, the detection of chaotic saddles and their stable and unstable manifolds. The numerical analysis is then presented in section \ref{section_analysis}. We begin by inspecting the parameter space and identifying a periodic window in the bifurcation diagram, then focusing on a saddle-node bifurcation and an interior crisis. Finally, the conclusion and the relevance of our results for the study of modulational processes in plasmas and fluids is presented in section \ref{section_conclusion}.

  \section{Nonlinear model} \label{section_model}

The nonlinear interaction between high-frequency Langmuir waves and low-frequency ion-acoustic waves in plasmas is described by the Zakharov equations \cite{nicholson:1983}

  \begin{eqnarray}
    \frac{\partial^2 \mathbf{E}_h}{\partial t^2} + \omega_{pe}^2 \mathbf{E}_h - \gamma_e v_{th}^2 \nabla^2 \mathbf{E}_h & = & - \omega_{pe}^2 \frac{n_1}{n_0}\mathbf{E}_h \label{eq_zakharov_1},\\
    \frac{\partial^2 n_1}{\partial t^2} - v_s^2 \nabla^2 n_1 & = & \frac{1}{2}\frac{\epsilon_0}{m_i} \nabla^2 \left< |\mathbf{E}_h|^2 \right>, \label{eq_zakharov_2}
  \end{eqnarray}

\noindent where $\mathbf{E}_h$ represents the high-frequency electric field fluctuations, $\omega_{pe}$ is the electron plasma frequency, $\gamma_e$ is the ratio of the specific heats for electrons, $v_{th}$ is the electron thermal velocity, $n_0$ is the average ion density, $n_1$ represents the ion density fluctuations (i.e., the deviations of the ion density from $n_0$), $v_s^2$ is the ion-acoustic velocity, $\epsilon_0$ is the permittivity of vacuum, $m_i$ is the ion mass, and the angular brackets represent the fast-scale time average (i.e., the time average performed in a period corresponding to the high-frequency fluctuations). Assuming that the Langmuir waves propagate along the $x$ axis we can write the electric field in the amplitude-modulated representation as

  \begin{displaymath}
    E_h (x, t) = \frac{1}{2} \hat{E}(x, t) \exp(-i\omega_{pe} t) + c. c.,
  \end{displaymath}

\noindent where $\hat{E}$ represents a slowly-varying complex envelope such that $|\partial_t^2 \hat{E}| \ll |\omega_{pe} \partial_t \hat{E}|$, $|\partial_x^2 \hat{E}| \ll |k \partial_x \hat{E}|$, and $i = \sqrt{-1}$. By defining the following dimensionless variables

  \begin{eqnarray*}
    z & = & \left( \frac{2}{\gamma_e} \right) \left( \frac{\eta m_e}{m_i} \right)^{1/2} \left(\frac{x}{\lambda_D} \right), \\
    \tau & = & \left( \frac{2}{\gamma_e} \right) \left( \frac{\eta m_e}{m_i} \right) \omega_{pe} t, \\
    E & = & \left(\frac{1}{\eta} \right) \left( \frac{m_i}{m_e} \right) ^{1/2} \left( \frac{\gamma_e \epsilon_0 \hat{E}^2}{8 n_0 K_B T_e} \right)^{1/2}, \\
    n & = & \left(\frac{\gamma_e}{4} \right) \left( \frac{m_i}{m_e \eta} \right) \left( \frac{n_{1}}{n_0} \right),
  \end{eqnarray*}

\noindent where $\eta = (\gamma_e K_B T_e + \gamma_i K_B T_i)/(K_B T_e)$, $K_B$ represents Boltzmann's constant, $T_e$ and $T_i$ are the electron and ion temperatures respectively, $\gamma_i$ the ratio of specific heats for ions, $m_e$ the electron mass, and $\lambda_D$ represents the Debye length, we can write Eqs. (\ref{eq_zakharov_1})-(\ref{eq_zakharov_2}) as

  \begin{eqnarray}
    i \frac{\partial E}{\partial \tau} + \frac{\partial^2 E}{\partial z^2} & = & n E \label{eq_normalized_zakharov_1},\\
    \frac{\partial^2 n}{\partial \tau^2} - \frac{\partial^2 n}{\partial z^2} & = & \frac{\partial |E|^2}{\partial z^2}. \label{eq_normalized_zakharov_2}
  \end{eqnarray}

Next, we make the approximation $\partial n/\partial \tau = 0$ in Eq. (\ref{eq_normalized_zakharov_2}), which can be justified as follows. Random fluctuations of the high-frequency electric field will produce undulations or ``ripples'' of Langmuir waves. These ripples imply a small gradient of wave intensity generating a force called the ponderomotive force, which moves electrons and ions towards regions where the intensity of the electric field is minimum. As a result, the plasma density will increase within these wave troughs. The dispersion relation of Langmuir waves

  \begin{displaymath}
    \omega_L^2 = \omega_{pe}^2 + \frac{3}{2} v_{th}^2 k_L^2,
  \end{displaymath}

\noindent indicates that, for a given frequency $\omega_L$, large $k_L$ wavenumbers can only exist in regions where $\omega_{pe}$ is small. Hence, Langmuir waves will be ``trapped'' in low-density regions. This localization of Langmuir waves allows the wave envelope to develop a growing ripple, which enhances the effect of the ponderomotive force \cite{chen:1984}. In the frame of reference of the high-frequency Langmuir wave, the low-frequency density perturbation becomes practically static. Hence, we can assume that the time derivative in Eq. (\ref{eq_normalized_zakharov_2}) vanishes.

Introducing the static approximation $\partial n/\partial \tau = 0$ in Eq. (\ref{eq_normalized_zakharov_2}), integrating twice and setting the constants of integration equal to zero yield

  \begin{equation}
    n = - |E|^2. \label{eq_density_electricfield}
  \end{equation}

\noindent By substituting Eq. (\ref{eq_density_electricfield}) into Eq. (\ref{eq_normalized_zakharov_1}) we obtain

  \begin{equation}
    i\frac{\partial E}{\partial \tau} + \frac{\partial^2 E}{\partial z^2} + |E|^2 E = 0, \label{eq_schrodinger_1}.
  \end{equation}

\noindent Equation (\ref{eq_schrodinger_1}) is called the Nonlinear Schr\"odinger equation because it resembles the Schr\"odinger equation of quantum mechanics \cite{nicholson:1983}. Following \citet{russell_ott:1981} we include the term $i\hat{\nu} E$ in Eq. (\ref{eq_schrodinger_1}), where $\hat{\nu}$ represents the linear growth/damping rate of $E$, and obtain

  \begin{equation} \label{eq_schrodinger_2}
    i\left( \frac{\partial E}{\partial t} + \hat{\nu} E \right) + \frac{\partial^2 E}{\partial z^2} + |E|^2 E = 0
  \end{equation}

\noindent We look for solutions of Eq. (\ref{eq_schrodinger_2}) in the form of three travelling waves

  \begin{eqnarray}
    E(z, \tau) & = & E_0(\tau) \exp[-i(k_0 z - \omega_0 \tau)] \nonumber \\
      & & + E_1(\tau) \exp[-i(k_1 z - \omega_1\tau)] \nonumber \\
      & & + E_2(\tau) \exp[-i(k_2 z - \omega_2\tau)], \label{eq_travelling_waves}
  \end{eqnarray}

\noindent where the resonant condition $2k_0 = k_1 + k_2$ and the linear dispersion $\omega_\sigma = k_\sigma^2$ ($\sigma = 0, 1, 2$) are assumed. Introducing (\ref{eq_travelling_waves}) in (\ref{eq_schrodinger_2}) we obtain the following set of complex ordinary differential equations (ODEs)

  \begin{eqnarray}
    \dot{E}_0 & = & -\hat{\nu}(k_0) E_0 + i[|E_0|^2 E_0 + 2 |E_1|^2 E_0 \nonumber \\
    & & + 2|E_0|^2 E_0 + 2 \bar{E}_0 E_1 E_2 \exp(2i\delta \tau)], \label{eq_model_complex1} \\
    \dot{E}_1 & = & -\hat{\nu}(k_1) E_1 + i[|E_1|^2 E_1 + 2 |E_0|^2 E_1 \nonumber \\
    & & + 2|E_2|^2 E_1 + E_0^2 \bar{E}_2 \exp(-2i\delta \tau)], \label{eq_model_complex2} \\
    \dot{E}_2 & = & -\hat{\nu}(k_2) E_2 + i[|E_2|^2 E_2 + 2 |E_0|^2 E_2 \nonumber \\
    & & + 2|E_1|^2 E_2 + E_0^2 \bar{E}_1 \exp(-2i\delta \tau)], \label{eq_model_complex3}
  \end{eqnarray}

\noindent where $\delta = (\delta_1 + \delta_2)/2$, $\delta_\sigma = \omega_\sigma - \omega_0$ $(\sigma = 1, 2)$ and $(\dot{})$ = $d/d\tau$. In terms of quantum mechanics, Eqs. (\ref{eq_model_complex1})-(\ref{eq_model_complex3}) denote the nonlinear temporal evolution of a 4-quanta system wherein a pair of pump quanta interact with a pair of Stokes and anti-Stokes daughter quanta. This set of equations can be rewritten using the polar representation $E_{\sigma} = a_{\sigma} (t) \exp[i \psi_{\sigma}(t)]$ $(\sigma = 0, 1, 2)$, where $a$ and $\psi$ are real amplitude and phase respectively, to obtain a set of four real ODEs

  \begin{eqnarray}
    \dot{a}_0 & = & \nu_0 a_0 + 2a_0 a_1 a_2 \sin\phi, \label{eq_model1} \\
    \dot{a}_1 & = & - \nu_1 a_1 - a_0^2 a_2 \sin\phi, \label{eq_model2} \\
    \dot{a}_2 & = & - \nu_2 a_2 - a_0^2 a_1 \sin\phi, \label{eq_model3} \\
    \dot{\phi} & = & - 2\delta + a_1^2 + a_2^2 - 2a_0^2 \nonumber \\
    & & + \left[ 4a_1 a_2 - a_0^2 \left( \frac{a_2}{a_1} + \frac{a_1}{a_2} \right) \right] \cos\phi, \label{eq_model4}
  \end{eqnarray}

\noindent where $\phi(t) = 2\psi_0 - \psi_1 - \psi_2 - 2\delta t$, $\nu_0 = -\hat{\nu}(k_0)$, $\nu_{1, 2} = \hat{\nu}(k_{1, 2})$. Wave $0$ represents a linearly unstable (pump) wave, and waves $1$ and $2$ represent linearly damped (daughter) waves. By assuming equally damping rates for the daughter waves ($\nu_2 = \nu_1 = \nu$) one can obtain $a_1 = a_2$, reducing Eqs. (\ref{eq_model1})-(\ref{eq_model4}) to a set of three ODEs

  \begin{eqnarray}
    \dot{a}_0 & = & \nu_0 a_0 + 2a_0 a_1^2 \sin\phi, \label{eq_simplified_1} \\
    \dot{a}_1 & = & -\nu a_1 - a_0^2 a_1 \sin\phi, \\
    \dot{\phi} & = & -2 \delta + 2 (a_1^2 - a_0^2) + 2(2a_1^2 - a_0^2) \cos\phi,\label{eq_simplified_3}
  \end{eqnarray}

\noindent which has been extensively studied \cite{russell_ott:1981, gosh_papadopoulos:1987,miranda_etal:2005}. In this paper we avoid this simplification and let $\nu_1 \ne \nu_2$ , retaining the four-dimensional set of equations (\ref{eq_model1})-(\ref{eq_model4}). Numerical integration of Eqs. (\ref{eq_model1})-(\ref{eq_model4}) is performed using the {\tt lsoda} package which is a variable-step integrator \cite{petzold:1983}. In this paper we define a Poincar\'e surface of section as $a_2 = 1$, and refer to ``Poincar\'e points'' as the intersection between the trajectory and the Poincar\'e surface with $\dot{a}_2 > 0$.

\section{Numerical algorithms} \label{section_algorithms}

\subsection{Computing Lyapunov exponents} \label{sec_lyapunov}

  The dynamical stability of a nonlinear system represented by a set of ODEs can be characterized by the Lyapunov exponents, which represent the exponential divergence or convergence rates of nearby trajectories \cite{oseledec:1968, shimada_nagashima:1979,benettin_etal:1980}. Let us rewrite Eqs. (\ref{eq_model1})-(\ref{eq_model4}) in the form

  \begin{equation} \label{eq_dynamic_form}
    \frac{d\mathbf{u}}{dt} = \mathbf{F}(\mathbf{u}),
  \end{equation}

\noindent where $\mathbf{u} = (a_0, a_1, a_2, \phi)$. Denote by $\mathbf{u}_0$ an initial condition of system (\ref{eq_dynamic_form}) at $t = t_0$, and $\theta_t (\mathbf{u}_0, t_0)$ its solution, that is

  \begin{equation} \label{eq_trajectory}
    \dot{\theta}_t (\mathbf{u}_0, t_0) = \mathbf{F}(\theta_t(\mathbf{u}_0, t_0), t),
  \end{equation}

\noindent and $\theta_{t_0}(\mathbf{u}_0, t_0) = \mathbf{u}_0$. Taking the derivative of Eq. (\ref{eq_trajectory}) with respect to $\mathbf{u}_0$ and defining the flux Jacobian $\Theta_t (\mathbf{u}_0, t_0) = D_{\mathbf{u}_0} \theta_t (\mathbf{u}_0, t_0)$ we obtain

  \begin{equation}
    \dot{\Theta}_t(\mathbf{u}_0, t_0) = D_{\mathbf{u}} \mathbf{F} ( \theta_t (\mathbf{u}_0, t_0), t) \Theta_t (\mathbf{u}_0, t_0), \label{eq_variational1}
  \end{equation}

\noindent and $\Theta_{t_0} (\mathbf{u}_0, t_0) = I$, where $I$ represents the identity matrix. Equation (\ref{eq_variational1}) is known as the variational equation \cite{steeb:1980,steeb_louw:1986}. A small perturbation $\delta \mathbf{u}_0$ of $\mathbf{u}_0$ will evolve as

  \begin{equation}
    \delta \mathbf{u} = \Theta_t(\mathbf{u}_0, t_0) \delta \mathbf{u}_0.
  \end{equation}

\noindent Suppose that $\delta \mathbf{u}$ can be decomposed into a linear combination of $j$ orthogonal vectors

  \begin{equation} \label{eq_decompos}
    \delta \mathbf{u} = \delta \mathbf{u}_1 + \delta \mathbf{u}_2 + ... + \delta \mathbf{u}_j, \qquad j = 1, ..., D,
  \end{equation}

\noindent where $D$ is the dimension of the phase space, and $\delta \mathbf{u}_0 = \delta \mathbf{u}_1(t_0) + \delta \mathbf{u}_2(t_0) + ... + \delta \mathbf{u}_j(t_0)$. The asymptotic behavior of perturbation $\delta \mathbf{u}$ is given by the Lyapunov spectrum, which is the set of Lyapunov exponents $\lambda_j$ defined by \cite{ott:1993}

  \begin{equation} \label{eq_lambdas}
    \lambda_j = \lim_{t \rightarrow \infty} \frac{1}{t - t_0} \ln \left| \frac{m_j(t)}{m_j(t_0)} \right|, \qquad j = 1, ..., D.
  \end{equation}

\noindent where $m_j$ are the eigenvalues of the flux Jacobian $\Theta_t$, obtained by solving Eq. (\ref{eq_variational1}). A positive Lyapunov exponent indicates that the distance between two trajectories will increase exponentially in time, whereas a negative value indicates that the distance will decrease exponentially in time. The number of Lyapunov exponents is equal to the dimension of the phase space, and each exponent represents the exponential divergence/convergence rate of initial conditions along a different direction in the phase space. The maximum Lyapunov exponent represents the rate in the direction of maximum divergence, or minimum convergence. Note that, in the case of continuous-time nonlinear models such as Eqs. (\ref{eq_model1})-(\ref{eq_model4}), at least one exponent is always equal to zero and represents the dynamics in the direction tangent to the flow, except when the asymptotic solutions converge to a fixed point. In this paper we will refer to $\lambda_{\mathrm{MAX}}$ as the maximum non-zero Lyapunov exponent. Chaos occurs when $\lambda_{\mathrm{MAX}} > 0$. If $\lambda_{\mathrm{MAX}} < 0$, then the dynamics can be periodic or quasiperiodic. Periodic behavior occurs when the Lyapunov spectrum is characterized by only one vanishing Lyapunov exponent and $\lambda_{\mathrm{MAX}} < 0$, whereas quasiperiodic behavior occurs when there are two or more vanishing exponents and $\lambda_{\mathrm{MAX}} < 0$.

  The numerical techniques for computing the Lyapunov spectrum of ODEs were first developed by \citet{shimada_nagashima:1979} and \citet{benettin_etal:1980} . In this paper we compute the Lyapunov spectrum of Equations (\ref{eq_model1})-(\ref{eq_model4}) following the numerical implementation found in Appendix A of \citet{wolf_etal:1985}.

  \subsection{Detecting chaotic saddles} \label{sec_chaotic_saddles}

    Since chaotic saddles are non-attracting chaotic sets, the trajectory of any initial condition arbitrarily close to the chaotic saddle will eventually leave its vicinity in forward-time dynamics, except for initial conditions located exactly in the chaotic saddle or its stable manifold. The stable manifold is defined as the set of points which converge to the chaotic saddle in forward-time dynamics, whereas the unstable manifold is the set of points which converge to the chaotic saddle in reverse-time dynamics \cite{hsu_etal:1988, nusse_yorke:1989}. We implement the following two methods to numerically find a chaotic saddle. 

  \subsubsection{The PIM-triple method}

    This method, proposed by \citet{nusse_yorke:1989}, defines a ``Proper Interior Maximum triple'', or PIM-triple, as the triple $(a, c, b)$ which lies on the line segment $(a, b)$ intersecting (or ``straddling'') the stable manifold of a chaotic saddle in phase space in such a way that point $c$ is closer to the stable manifold than $a$ and $b$, and the escape time profile of the triple $(a, c, b)$ will have a maximum at $c$. This procedure can be described as follows

  \begin{enumerate}
    \item Define a region of the phase space with no attractors, called the restraining region. Define the escape time as the time it takes for a trajectory to leave the restraining region. Select two points $a$ and $b$ on the Poincar\'e surface which straddle the stable manifold of the chaotic saddle in phase space, and set the error $\epsilon$ to a small value.

    \item \label{item_pimrefine} Divide the segment $(a, b)$ into $M$ subsegments, and compute the escape time of each subsegment. Choose one subsegment $(a_1, c_1, b_1)$ which qualifies as a PIM-triple.

    \item Repeat step \ref{item_pimrefine} until the Euclidean length of the segment $(a_n, b_n)$ is smaller than $\epsilon$.

    \item Iterate the Poincar\'e map for initial conditions $a_n$ and $b_n$, obtaining a new pair $(a_n, b_n)$. Repeat this step until the length of the segment $(a_n, b_n)$ is larger than $\epsilon$.

    \item Repeat step \ref{item_pimrefine}.
  \end{enumerate}
  This method allows one to obtain a numerical trajectory which will stay very close to the chaotic saddle for an arbitrarily long period of time within an error $\epsilon$, and it is useful to compute statistical quantities which require long trajectories, such as the Lyapunov exponents. Its main disadvantage is that the PIM-Triple method may not work in the presence of two or more positive Lyapunov exponents \cite{nusse_yorke:1989} (i.e., the stable manifold of the chaotic saddle has codimension higher than one). Instead, one can use other alternatives such as the PIM-simplex algorithm \cite{moresco_dawson:1999}, the Stagger-and-Step method \cite{sweet_etal:2001}, or the gradient search algorithm \cite{bollt:2005}. Our implementation of the PIM-triple method follows the original description by \citet{nusse_yorke:1989}, and we set $\epsilon = 10^{-6}$.

  \subsubsection{The sprinkler method}

  In the sprinkler method \cite{kantz_grassberger:1985,hsu_etal:1988} the restraining region in the Poincar\'e surface of section is covered by a two-dimensional (2D) grid of initial conditions, and then each initial condition is iterated until some time $t_c$, which should be larger than the average escape time from the restraining region, and must be adjusted after some trial and error. The final points in the Poincar\'e surface that remain in the restraining region after $t_c$ time units approximate the unstable manifold, their initial conditions approximate the stable manifold, and the points obtained at $\bar{t} = \xi t_c$ approximate the chaotic saddle. For most systems (including Eqs. (\ref{eq_model1})-(\ref{eq_model4})) $\xi = 0.5$ works fine \cite{hsu_etal:1988,rempel_etal:2004a}. The accuracy of this method can be improved by choosing a large $t_c$, however, the number of initial conditions in the grid must be increased as well. Our implementation of this method is based on the description given by \citet{hsu_etal:1988}, using a grid of $1000 \times 1000$ initial conditions, and we set $t_c = 20$ time units.

  The chaotic saddle must coincide visually with the crossings of its stable and unstable manifolds obtained using the Sprinkler method. Since in our case the Poincar\'e surface is a three-dimensional hypersurface, the visualization of the chaotic saddle and its stable and unstable manifolds can be affected by the choice of the 2D grid of initial conditions. We overcome this by choosing a grid of initial conditions in $(a_0, a_1)$, and setting the values of $\phi$ to $\phi = P(a_0)$, where $P$ is a second-order polynomial which was obtained by plotting $\phi$ as a function of $a_0$ and computing a quadratic fit using the Poincar\'e points of the chaotic saddle. This ensures that the chaotic saddle and its manifolds are entirely contained in the surface defined by $(a_0, a_1, P(a_0)$), and the intersections between the stable and unstable manifolds coincide with the chaotic saddle. Details on this procedure will be provided in a future work \cite{miranda_etal:2011}.

\section{Nonlinear analysis} \label{section_analysis}

\subsection{Parameter space}

\begin{figure}

  \begin{center}
    \includegraphics[width=0.7\textwidth]{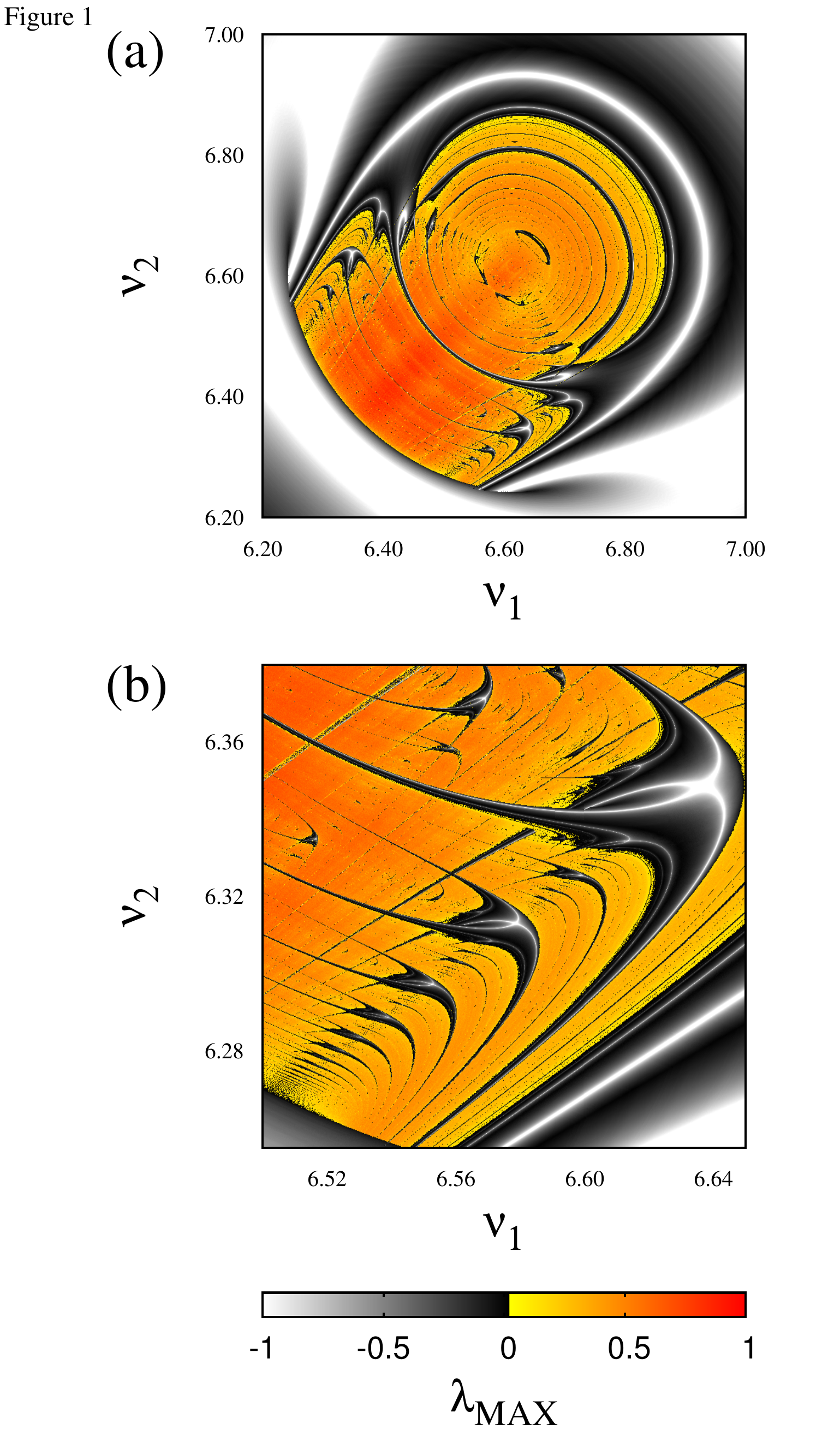}
  \end{center}

  \caption{(Color online) Shrimp-shaped self-similar structures in a two-parameter space. Upper panel: Maximum non-zero Lyapunov exponent $\lambda_{\mathrm{MAX}}$ as a function of $\nu_1$ and $\nu_2$. The colour scale indicates chaotic dynamics, whereas the gray scale denotes periodic oscillations. Lower panel: Enlargement of a subregion of (a).}
  \label{fig0}
\end{figure}

  We begin our numerical analysis by fixing $\delta = -6$ following Refs. \onlinecite{russell_ott:1981, miranda_etal:2005}, and inspecting the stability of asymptotic solutions of Eqs. (\ref{eq_model1})-(\ref{eq_model4}) as a function of the damping rates $\nu_1$ and $\nu_2$, i.e., in a two-dimensional section of the parameter space. Figure \ref{fig0}(a) shows the value of the maximum non-zero Lyapunov exponent $\lambda_{\mathrm{MAX}}$ in parameter space. This figure was obtained as follows

  \begin{enumerate}
    \item \label{item_step1} Define a region of the parameter space by setting starting and ending values for $\nu_1$ and $\nu_2$. 

    \item \label{item_initconds} Set initial conditions and solve Eqs. (\ref{eq_model1})-(\ref{eq_model4}) until $t = T$ to discard transient values.

    \item \label{item_wolf} Obtain $\lambda_{\mathrm{MAX}}$ by computing the Lyapunov spectrum as explained in Section \ref{sec_lyapunov}.

    \item After $t = T_{\mathrm{MAX}}$, enough to ensure the convergence of $\lambda_{\mathrm{MAX}}$, increase the value of $\nu_2$ by a small amount $\Delta \nu_2$ and use the last values of the integrated trajectory as initial conditions of Eqs. (\ref{eq_model1})-(\ref{eq_model4}), to ensure that the same attractor is being ``followed'' across the parameter space. Repeat step \ref{item_wolf}.

    \item After reaching the ending value of $\nu_2$, reset $\nu_2$ to its starting value, increase $\nu_1$ by a small amount $\Delta \nu_1$, and repeat from step \ref{item_initconds}.
  \end{enumerate}

\noindent In our case we set $(\nu_1, \nu_2) \in [6.2,7]\times[6.2,7]$, $T = 3000$, $T_{\mathrm{MAX}} = 20000$, and $\Delta \nu_1 = \Delta \nu_2 = 10^{-3}$. Positive values of $\lambda_{\mathrm{MAX}}$ are represented using a colour scale, whereas negative values are shown using a gray scale. The colour scale varies from yellow (smaller values) to red (larger values), whereas the gray scale varies from white (smaller values) to black (larger values). Figure \ref{fig0}(a) clearly displays a central area of chaotic dynamics ($\lambda_{\mathrm{MAX}} > 0$) surrounded by ordered dynamics ($\lambda_{\mathrm{MAX}} < 0$). Embedded in the chaotic area are regular structures, enclosing periodic dynamics, identified as ``shrimps'' with a characteristic shape consisting of a main body and four elongated ``antennae''. Fig. \ref{fig0}(b) shows a detailed view of these self-similar structures, revealing a complex network of paths, or ``window streets'' \cite{lorenz:2008} composed by the entangled antennae and their intersections interconnecting shrimps. These paths allow one to move through the parameter space, while keeping the dynamics periodic. It is clear that the rich structures found in the parameter space are symmetric with respect to the line $\nu_2 = \nu_1$, which corresponds to the dynamics of the reduced system (\ref{eq_simplified_1})-(\ref{eq_simplified_3}).

\subsection{Bifurcation diagram}

\begin{figure}

  \begin{center}
    \includegraphics[width=0.7\textwidth]{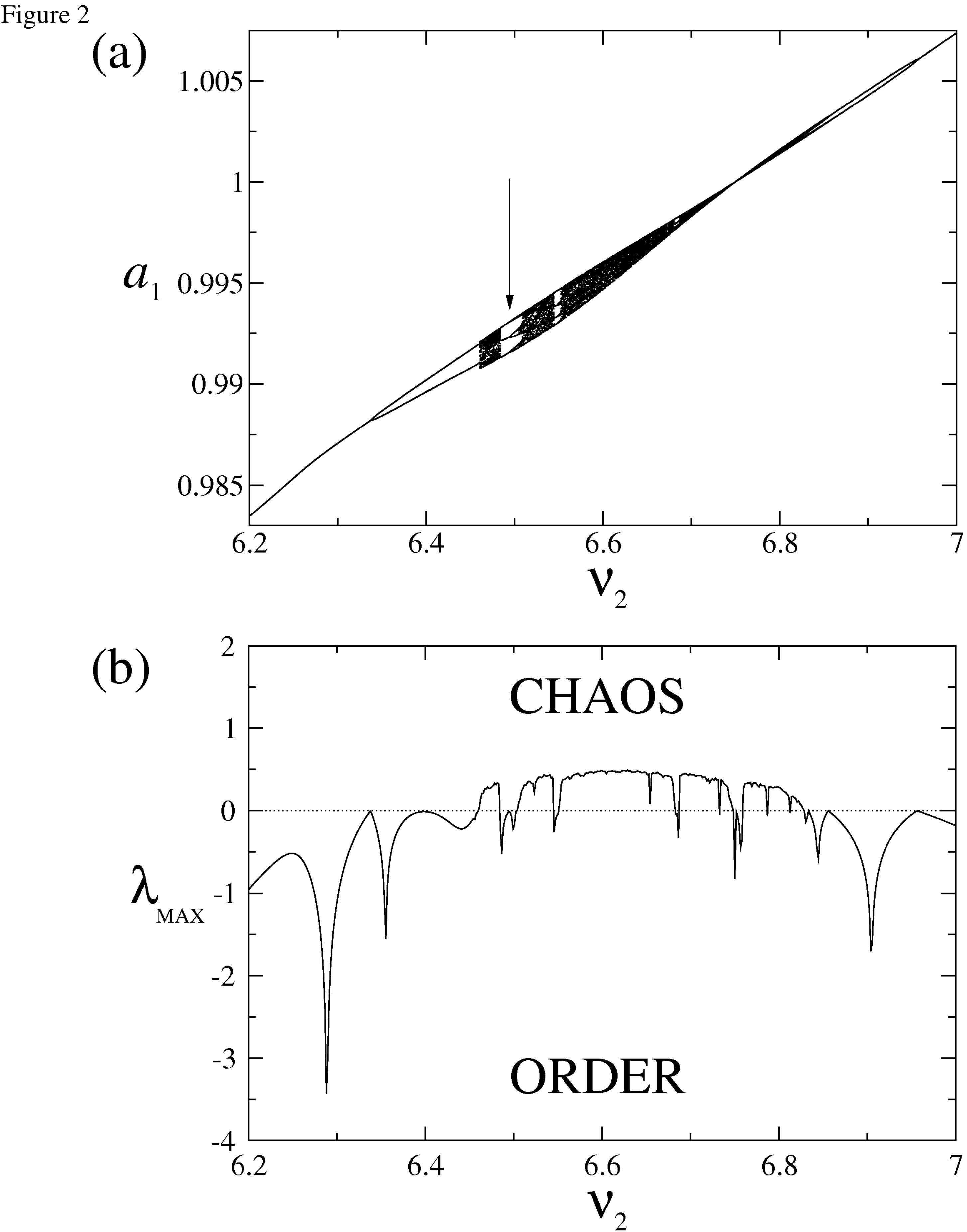}
  \end{center}

  \caption{(a) Bifurcation diagram for $a_1$ as a function of $\nu_2$. The arrow indicates a period-3 window. (b) Maximum non-zero Lyapunov exponent $\lambda_{\mathrm{MAX}}$ as a function of $\nu_2$.}
  \label{fig1}
\end{figure}

Next, we construct a bifurcation diagram by fixing $\nu_1 = 6.75$ following previous works \cite{russell_ott:1981, miranda_etal:2005}, and choosing $\nu_2$ as our control parameter. Figure \ref{fig1}(a) shows an overview of the bifurcation diagram using the amplitude of the first daughter wave $a_1$ as a function of $\nu_2$. This figure was constructed as follows. Set $\nu_2 = 6.2$, and set the initial conditions. Solve Eqs. (\ref{eq_model1})-(\ref{eq_model4}), discarding the first 1500 transient Poincar\'e points, and save the next 300 points, plotting $a_1$ against $\nu_2$. Then, increase $\nu_2$ by a small step, use the last Poincar\'e point as the initial condition, and solve again the evolution equations until $\nu_2 = 7$. The resulting bifurcation diagram shown in Fig. \ref{fig1}(a) agrees with the chaotic and periodic regimes seen in Fig. \ref{fig0}(a). From Fig. \ref{fig1}(a) we observe the presence of a limit cycle at $\nu_2 = 6.2$, represented by a period-1 attractor. This means that the nonlinear waves described by the model equations display periodic dynamics. By increasing the control parameter the period of the limit cycle is doubled at $\nu_2 \sim 6.34$. At $\nu_2 \sim 6.45$ there is a sudden transition to a chaotic regime, which is occasionally interrupted by periodic windows as $\nu_2$ is increased. The largest periodic window occurs at $\nu_2 \sim 6.5$, and is indicated by an arrow. The bifurcation diagram becomes very narrow with increasing $\nu_2$. At $\nu_2 \sim 6.85$ the solutions of the model equations converge asymptotically to a period-2 attractor. At $\nu_2 \sim 6.95$ the periodicity of the attractor is reduced to 1.

We compute the maximum non-zero Lyapunov exponent $\lambda_{\mathrm{MAX}}$ for the same range of $\nu_2$. Fig. \ref{fig1}(b) shows the value of $\lambda_{\mathrm{MAX}}$ as a function of $\nu_2$. This figure is in agreement with Fig. \ref{fig1}(a) in the sense that periodic solutions correspond to negative values of $\lambda_{\mathrm{MAX}}$ and the chaotic region is characterized by positive values of $\lambda_{\mathrm{MAX}}$. Periodic windows are reflected by sudden drops of the value of $\lambda_{\mathrm{MAX}}$ towards negative values. Negative values of $\lambda_{\mathrm{MAX}}$ for $\nu_2 > 6.85$ confirm that the system dynamics is periodic.

\begin{figure}

  \begin{center}
    \includegraphics[width=\textwidth]{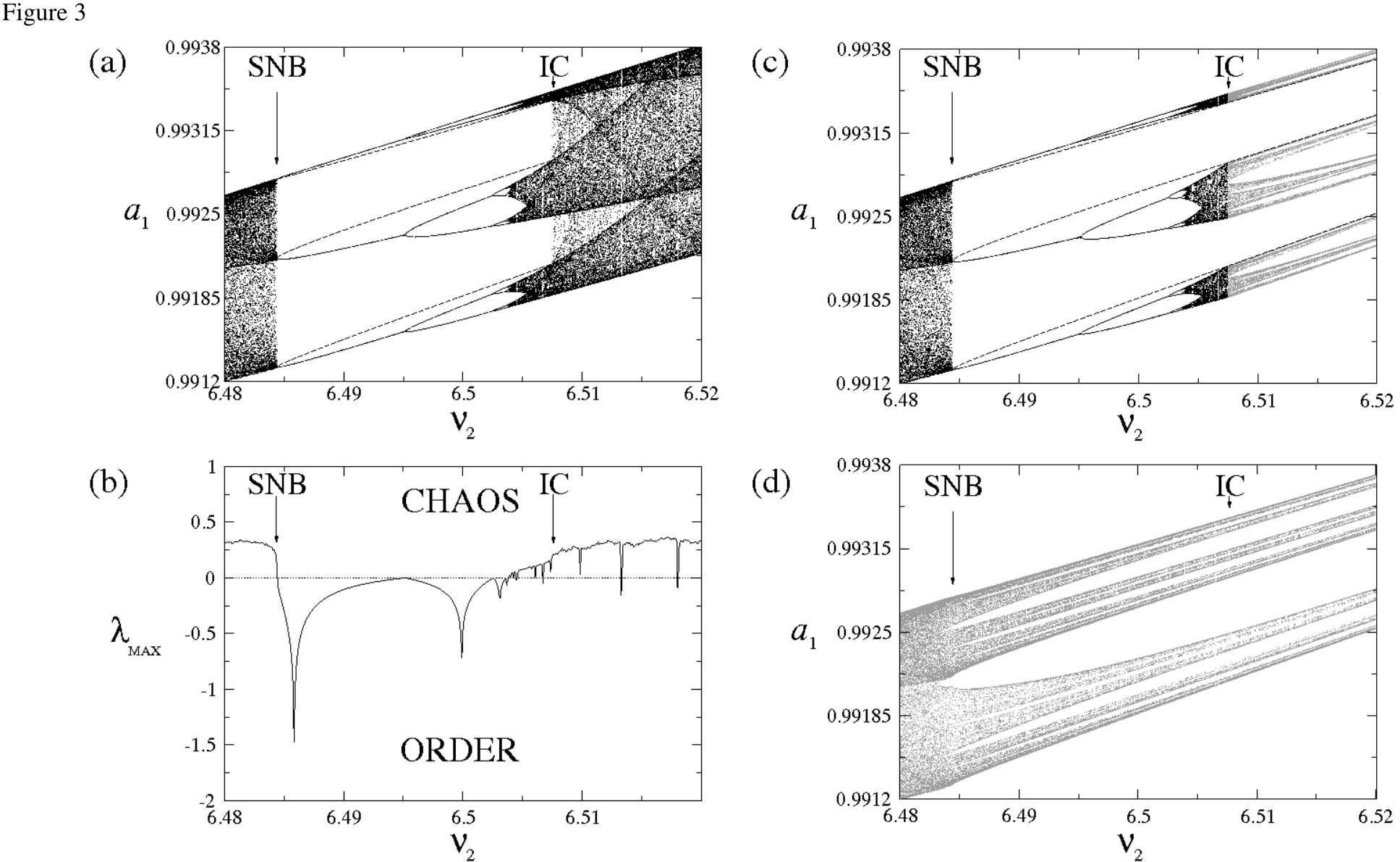}
  \end{center}

  \caption{(a) Bifurcation diagram for $a_1$ as a function of $\nu_2$. The dashed lines represent the evolution of the period-3 unstable periodic orbit, SNB denotes a saddle-node bifurcation and IC denotes an interior crisis. (b) Maximum non-zero Lyapunov exponent $\lambda_{\mathrm{MAX}}$ as a function of $\nu_2$. (c) Bifurcation diagram showing the conversion of the weak chaotic attractor (black) to a banded chaotic saddle (gray) and its evolution after crisis. (d) Bifurcation diagram showing the evolution of the surrounding chaotic saddle.}
  \label{fig2}
\end{figure}

We focus our analysis on the largest periodic window indicated by an arrow in Fig. \ref{fig1}, shown in detail in Fig. \ref{fig2}(a). This periodic window is created by a tangent bifurcation at $\nu_2 = \nu_{\mathrm{SNB}} \sim 6.4845$ known as a saddle-node bifurcation (SNB), where the chaotic attractor disappears and a stable and an unstable period-3 orbits are created. The stable periodic orbit (SPO) is displayed as a continuous line, while the unstable periodic orbit (UPO) is represented by a dashed line. The SPO undergoes a period-doubling cascade with increasing $\nu_2$, becoming a period-6 SPO at $\nu_2 \sim 6.495$, and a period-12 SPO at $\nu_2 \sim 6.5025$. Eventually, the periodicity of the SPO goes to infinity, becoming a chaotic attractor which can be distinguished as a set of three bands in Fig. \ref{fig2}(a). At $\nu_2  = \nu_{\mathrm{IC}} \sim 6.5074$ the chaotic attractor collides with the UPO created at the SNB and suddenly increases in size. This global bifurcation is known as an interior crises (IC), and marks the end of the periodic window. The UPO which mediates the IC is called a mediating UPO.

The computation of the maximum Lyapunov exponent shown in Fig. \ref{fig2}(b) clearly demonstrates the transition from chaotic to ordered states at SNB. At IC, there is a sudden increase in the value of $\lambda_{\mathrm{MAX}}$, for this reason the pre-crisis chaotic attractor with smaller $\lambda_{\mathrm{MAX}}$ is called a weak chaotic attractor (WCA), while the enlarged, post-crisis chaotic attractor with larger $\lambda_{\mathrm{MAX}}$ is called a strong chaotic attractor (SCA).

After colliding with the mediating UPO the WCA loses stability, becoming a nonattracting chaotic set confined to the same banded region. For this reason it is called a banded chaotic saddle (BCS). Fig. \ref{fig2}(c) shows the same bifurcation diagram as in Fig. \ref{fig2}(a), from $\nu_2 = 6.48$ to $\nu_2 = \nu_{\mathrm{IC}}$. After IC, we remove the SCA and plot the evolution of the BCS as a function of $\nu_2$. The BCS was obtained using the PIM-triple procedure, and the restraining region is defined as the same region previously occupied by the WCA, bounded by the mediating UPO and its stable manifold. Fig. \ref{fig2}(c) clearly shows the presence of empty regions or ``gaps'' that widen as $\nu_2$ increases.

Another type of chaotic saddle, which exists for the entire bifurcation diagram, is shown in Fig. \ref{fig2}(d). This chaotic saddle is called a surrounding chaotic saddle (SCS), because it ``surrounds'' the space occupied by the attractors within the periodic window, and the space occupied by the BCS after the interior crisis. The SCS was obtained using the PIM-triple procedure. The restraining region was defined in different ways, depending on the value of $\nu_2$. For $\nu_{\mathrm{SNB}} < \nu_2 < \nu_{\mathrm{IC}}$ (i.e., inside the periodic window), in the periodic regime, we cover the SPO with a small disc of radius $r = 10^{-7}$, and define the restraining region as the region of the phase space surrounding the SPO minus the disk. In the chaotic regime, the restraining region is defined as the region outside a rectangle that covers one of the bands of the WCA. For $\nu_2 > \nu_{\mathrm{IC}}$ (i.e., after the interior crisis) we define the restraining region as the region located outside that occupied by the BCS. For $\nu_2 < \nu_{\mathrm{SNB}}$ (i.e., before SNB) we use the same region as the first value of $\nu_2$ in the periodic regime. Similarly to the BCS shown in Fig. \ref{fig2}(c), the SCS displays ``gaps'' that become wider with increasing $\nu_2$, as displayed in Fig. \ref{fig2}(d).

\begin{figure}

  \begin{center}
    \includegraphics[width=\textwidth]{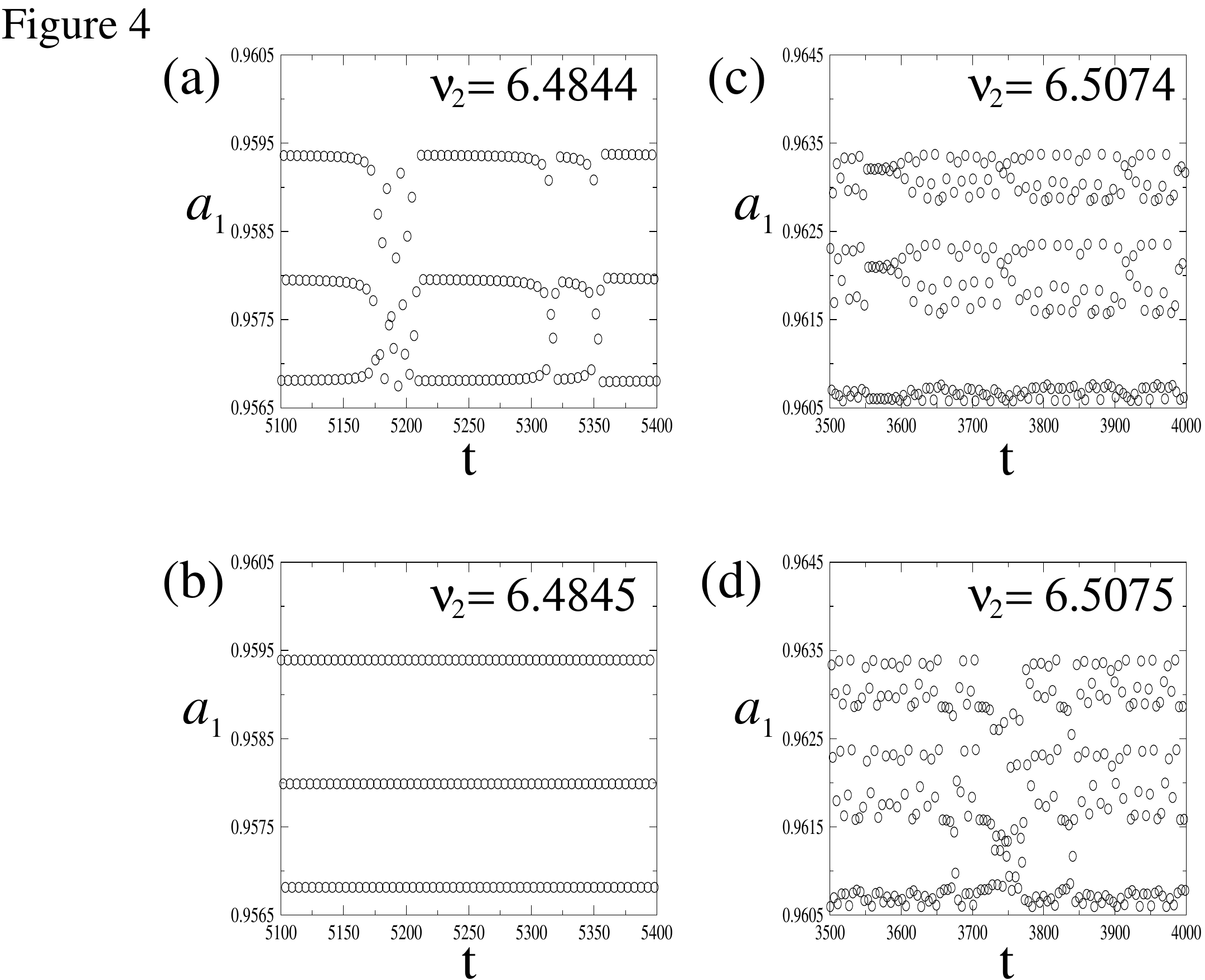}
  \end{center}

  \caption{(a) Time series of the Poincar\'e points of $a_1$ at $\nu_2 = 6.4844$, showing the characteristics of a type-I Pomeau-Manneville intermittency. (b) Time series of the Poincar\'e points of $a_1$ at $\nu_2 = 6.4845 > \nu_{\mathrm{SNB}}$, showing period-3 dynamics. (c) Time series of pre-crisis weak chaos represented by the Poincar\'e points of $a_1$ at $\nu_2 = 6.5074 \lesssim \nu_{\mathrm{IC}}$. (d) Time series of $a_1$ at $\nu_2 = 6.5075 \gtrsim \nu_{\mathrm{IC}}$ represented by the Poincar\'e points, showing a crisis-induced intermittency alternating episodically between periods of weak and strong chaos.}
  \label{fig3}
\end{figure}

The SNB, which marks the beginning of the periodic window shown in Fig. \ref{fig2}, is characterized by a sudden transition from chaotic to periodic dynamics. Before SNB, the chaotic time series displays type-I Pomeau-Manneville intermittency \cite{pomeau_manneville:1980} in which intervals resembling a periodic behavior are interrupted by occasional chaotic bursts. Figure \ref{fig3}(a) shows the time series of the first daughter wave $a_1$, represented by Poincar\'e points, for $\nu_2 = 6.4844 \lesssim \nu_{\mathrm{SNB}}$, which clearly displays intermittent behavior. After SNB, the asymptotic dynamics converges to a period-3 solution, as evidenced by the time series shown in Fig. \ref{fig3}(b) for $\nu_2 = 6.4845 \gtrsim \nu_{\mathrm{SNB}}$. Recall that the period-3 stable periodic orbit created at SNB undergoes a period-doubling cascade with increasing $\nu_2$, and that the periodicity of the SPO eventually tends to infinity, leading to the WCA. In this regime the time series of Poincar\'e points of the first daughter wave $a_1$ is characterized by weakly chaotic dynamics, and the points remain confined to three narrow bands as shown in Fig. \ref{fig3}(c) for $\nu_2 = 6.5074 \lesssim \nu_{\mathrm{IC}}$. After colliding with the mediating UPO at IC, which marks the end of the periodic window, the attractor increases in size and becomes the SCA. The time series of the daughter wave $a_1$ now displays strong ``bursts'' which occasionally interrupt ``laminar'' dynamics that keep the memory of the weak chaotic regime, as observed in Fig. \ref{fig3}(d) for $\nu_2 = 6.5075 \gtrsim \nu_{\mathrm{IC}}$. This intermittency is known as crisis-induced intermittency \cite{grebogi_etal:1983, grebogi_etal:1987}.

\begin{figure}

  \begin{center}
    \includegraphics[height=0.7\textheight]{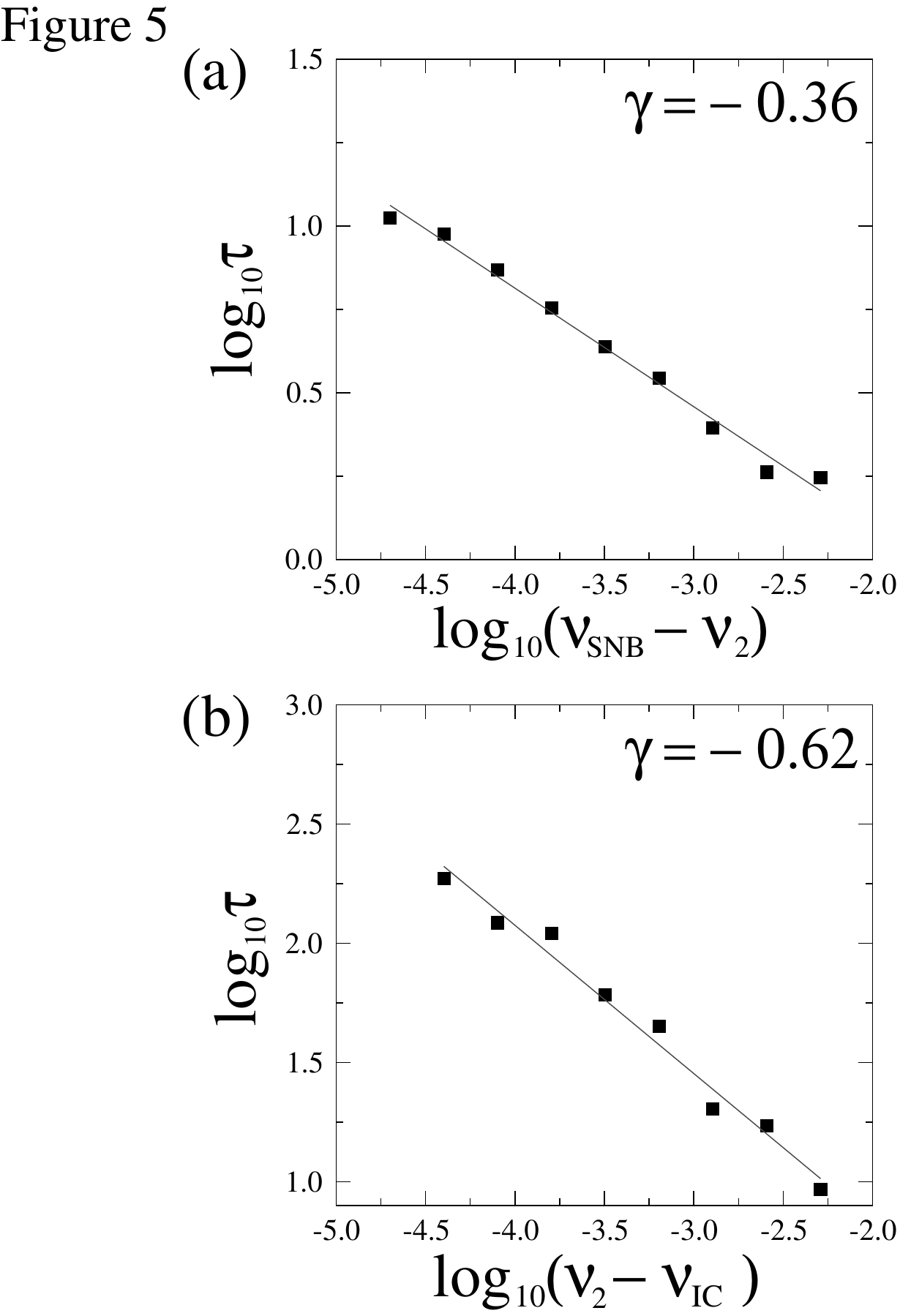}
  \end{center}

  \caption{(a) Average time $\tau$ between intermittent events (black squares) as a function of the departure of $\nu_2$ from $\nu_{\mathrm{SNB}}$ in the chaotic regime before the saddle-node bifurcation. The straight line represents a linear fit with a slope $\gamma = -0.36$. (b) Average time $\tau$ between intermittent bursts (black squares) as a function of the departure of $\nu_2$ from $\nu_{\mathrm{IC}}$ in the chaotic regime after the interior crisis, and a linear fit (straight line) with a slope $\gamma = -0.62$, in a log-log scale.}
  \label{fig4}
\end{figure}

The chaotic bursts observed in the time series of Figure \ref{fig3}(a), due to the type-I Pomeau-Manneville intermittency, become more frequent as the value of the control parameter departs from the SNB point, within the chaotic regime. This means that the average time $\tau$ of the intervals resembling periodic behavior decreases with decreasing $\nu_2$. Figure \ref{fig4}(a) shows $\tau$ as a function of the departure of $\nu_2$ from $\nu_{\mathrm{SNB}}$ (black squares) in log-log scale. This figure confirms that $\tau$ decreases as the ``distance'' between $\nu_2$ and $\nu_{\mathrm{SNB}}$ increases. Clearly, the average time follows a power-law, with a slope $\gamma = -0.36$ obtained by a linear fit represented by a continuous line in Fig. \ref{fig4}(a). Similarly, the average time $\tau$ in which the trajectory is confined to three bands as observed in the intermittent time series of Fig. \ref{fig3}(d) due to the crisis-induced intermittency decreases with the departure of $\nu_2$ from $\nu_{\mathrm{IC}}$ in the chaotic regime \cite{grebogi_etal:1987}, and follows a power-law behavior with a slope $\gamma = -0.62$, as shown in Fig. \ref{fig4}(b).

\subsection{Saddle-node bifurcation}

\begin{figure}

  \begin{center}
    \includegraphics[width=0.45\textwidth]{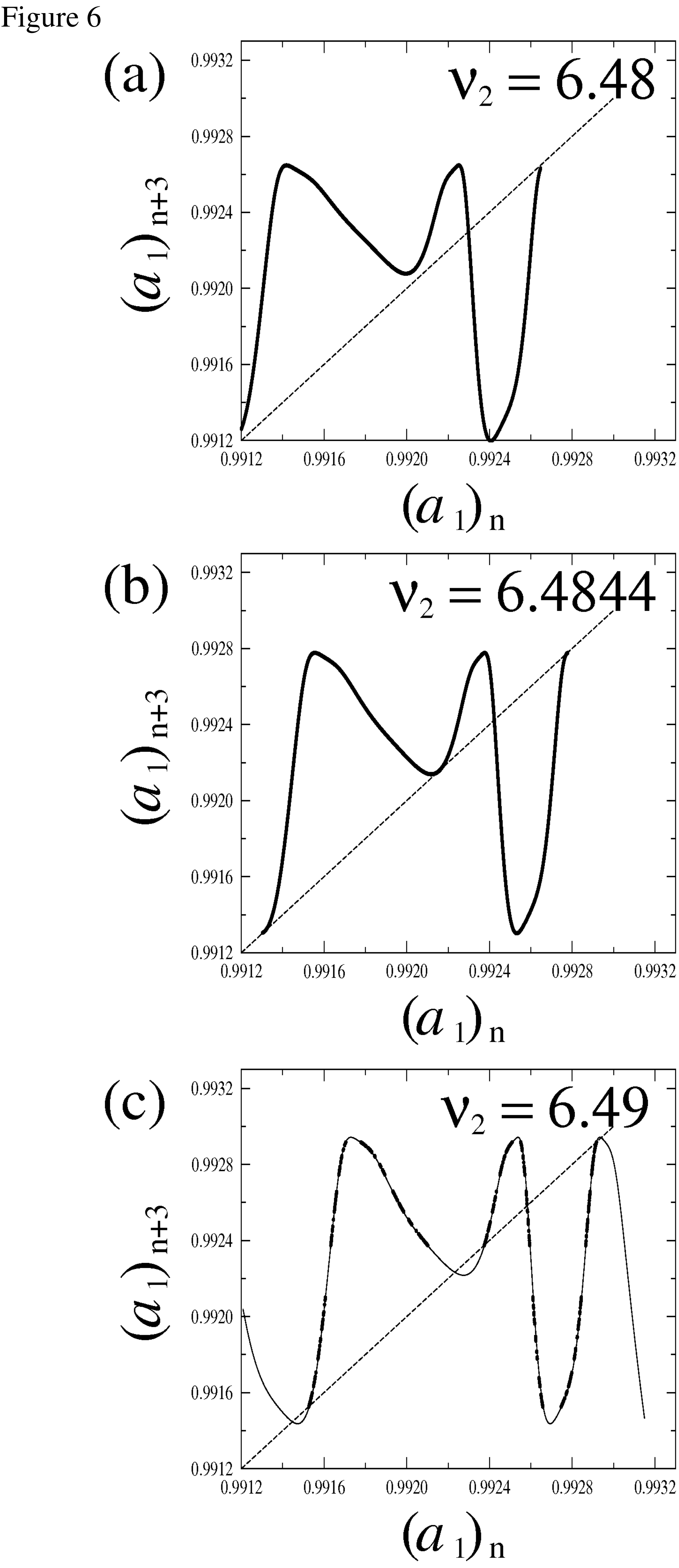}
  \end{center}

  \caption{Return map ($(a_1)_n$, $(a_1)_{n+3}$) for (a) a chaotic attractor at $\nu_2 = 6.48 < \nu_{\mathrm{SNB}}$. The dashed line represents $(a_1)_n = (a_1)_{n+3}$. (b) Chaotic attractor at $\nu_2 = 6.4844 \lesssim \nu_{\mathrm{SNB}}$. (c) Surrounding chaotic saddle at $\nu_2 = 6.49 > \nu_{\mathrm{SNB}}$ (black dots) and a curve obtained using a cubic fit (thin line).}
  \label{fig5}
\end{figure}

The SNB marks the beginning of the period-3 window shown in Fig. \ref{fig2}(a), where the chaotic attractor disappears and a stable periodic orbit and an unstable periodic orbit are created. In the Poincar\'e plane, periodic orbits are represented by fixed points. The creation of fixed points across SNB can be elucidated by constructing return maps. Figure \ref{fig5}(a) shows the value of the amplitude of the first daughter wave $a_1$ at the $n$-th intersection with the Poincar\'e map, denoted by $(a_1)_n$, plotted against its third crossing with the map, represented by $(a_1)_{n+3}$, using points from a chaotic trajectory at $\nu_2 = 6.48 < \nu_{\mathrm{SNB}}$ (thick line). Fixed points of periods 1 and 3 can be identified as intersections between the return map and the line given by $(a_1)_{n+3} = (a_1)_n$ (dashed line). From Fig. \ref{fig5}(a) the return map intersects the line at $(a_1)_n \sim 0.9923$. The return map also approaches the line at $(a_1)_n \sim 0.9912$, $(a_1)_n \sim 0.9920$, and at $(a_1)_n \sim 0.99265$. By increasing the value of the control parameter to $\nu_2 = 6.4844 \lesssim \nu_{\mathrm{SNB}}$ the return map becomes tangent to the line at the same three points (Fig. \ref{fig5}(b)). For $\nu_2 = 6.49 > \nu_{\mathrm{SNB}}$, since the dynamics of the attractor changed from chaotic to periodic we construct the return map using points from the surrounding chaotic saddle (SCS, black points). Due to its fractal nature the SCS has empty spaces or ``gaps'' which prevent us from demonstrating the crossings between the return map and the dashed line. We solve this by fitting the Poincar\'e points of the SCS to a curve using a cubic fit, and superposing the interpolated points against the return map. The return map obtained by this technique is shown as a thin line in Fig. \ref{fig5}(c). It is evident that the interpolated points fit well the return map of the SCS, and the intersections between the return map and the line $(a_1)_n = (a_1)_{n+3}$ clearly demonstrate the creation of a pair of stable-unstable period-3 fixed points.

\subsection{Interior crisis}

\begin{figure}

  \begin{center}
    \includegraphics[width=\textwidth]{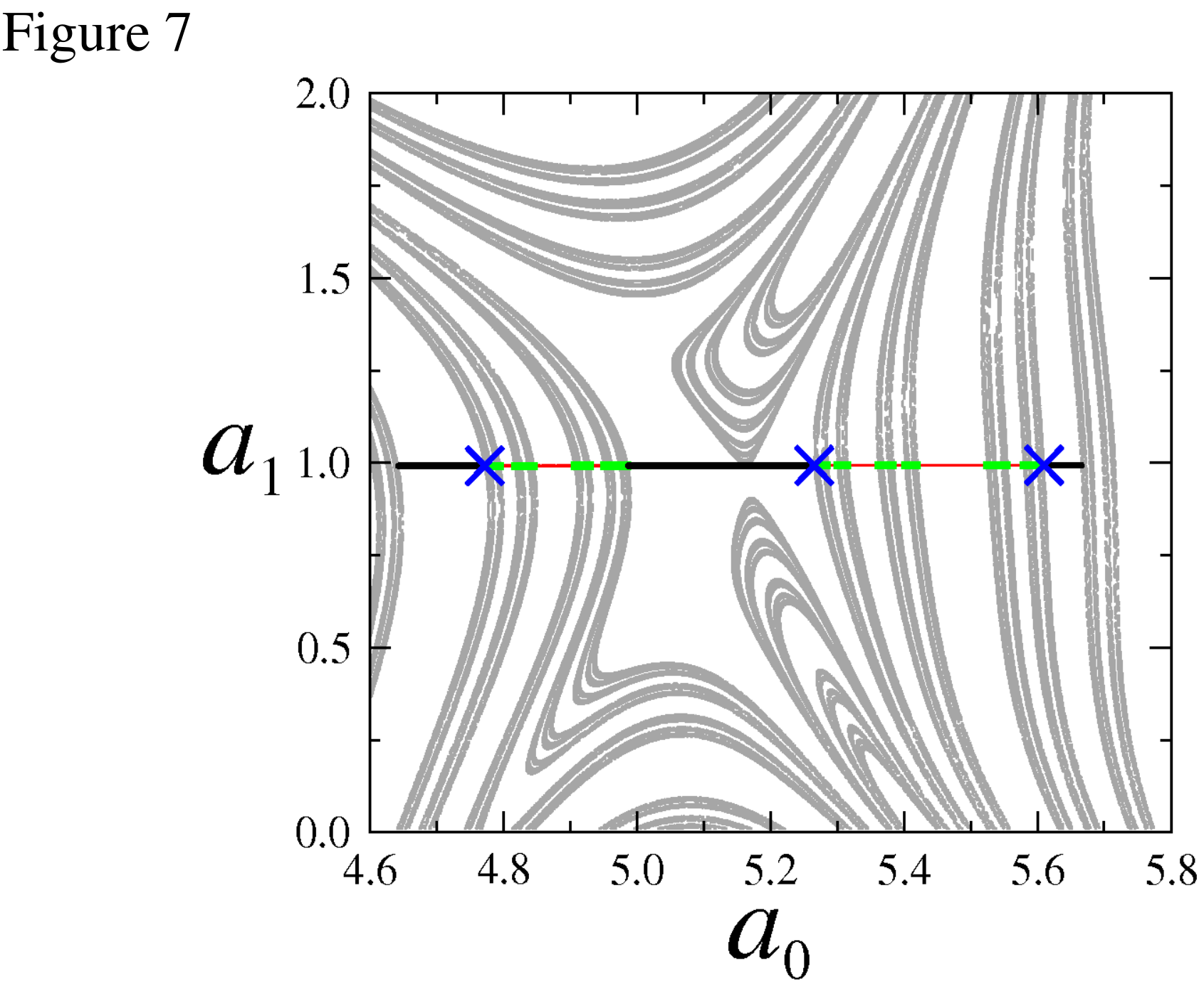}
  \end{center}

  \caption{(Color online) Poincar\'e plots of the surrounding chaotic saddle (green) right before crisis ($\nu_2 = 6.5074 \lesssim \nu_{\mathrm{IC}}$) and its stable (gray) and unstable (red) manifolds. The weak chaotic attractor is represented by black thick lines, and the period-3 mediating unstable periodic orbit by blue crosses.}
  \label{fig8}
\end{figure}

\begin{figure}

  \begin{center}
    \includegraphics[width=0.49\textwidth]{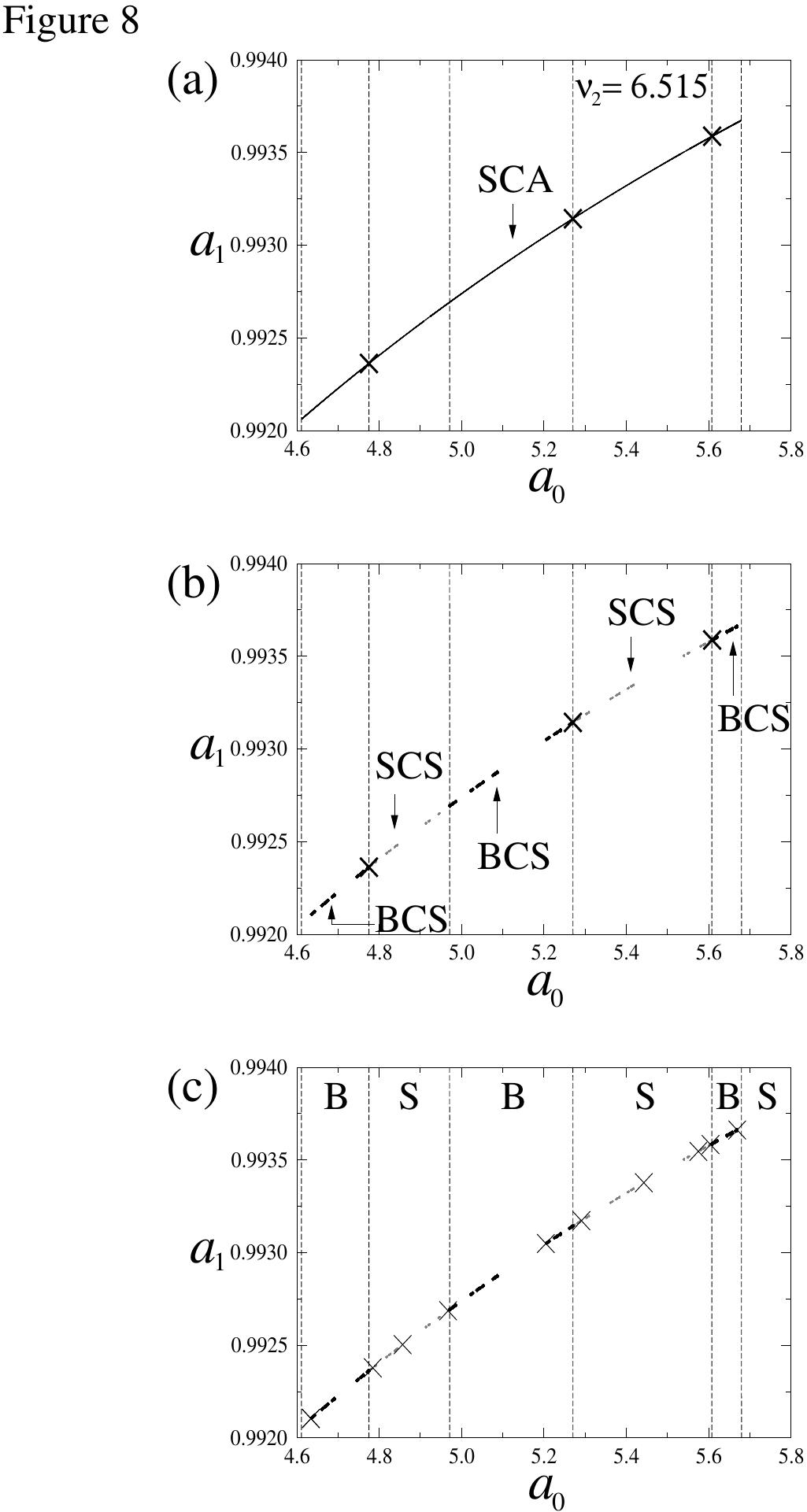}
  \end{center}

  \caption{Coexistence of a chaotic attractor, chaotic saddles and a coupling orbit after the interior crisis ($\nu_2 = 6.515 > \nu_{\mathrm{IC}}$). (a) Strong chaotic attractor (SCA) and the mediating unstable periodic orbit (UPO) represented by crosses. The dashed lines represent the stable manifold of the mediating UPO. (b) Banded chaotic saddle (BCS, black) and surrounding chaotic saddle (SCS, gray). (c) BCS and SCS superposed by a period-14 coupling unstable periodic orbit. The banded region is denoted by B and the surrounding region by S.}
  \label{fig9}
\end{figure}

The interior crisis is characterized by the collision between the WCA and the mediating UPO and its stable manifold at IC \cite{grebogi_etal:1983}. A common technique to compute the stable manifold of an UPO requires the integration of the model equations backwards in time \cite{you_etal:1991}. However, Equations (\ref{eq_model1})-(\ref{eq_model4}) turned out to be numerically unstable to be solved backwards in time with the traditional numerical integration schemes. Since the stable manifold of the mediating UPO and the stable manifold of the chaotic saddle are practically indistinguishable \cite{rempel_etal:2004a, ziemniak_etal:1994}, an alternative approach is to compute the stable manifold of the SCS and extract its ``contour'' to represent the stable manifold of the mediating UPO near the collision site. Figure \ref{fig8} shows an overview of the stable (gray) and unstable (red) manifolds of the SCS (green), for $\nu_2 = 6.5074 \lesssim \nu_{\mathrm{IC}}$. The WCA is shown in black, and the period-3 mediating UPO is represented by crosses. The SCS and its stable and unstable manifolds are obtained using the sprinkler method, in which a grid of $1000 \times 1000$ initial conditions in the $(a_0, a_1)$ plane was defined, and $\bar{t} = 10$ as indicated in Section \ref{sec_chaotic_saddles}. We use the following second-degree polynomial obtained by a quadratic fit to set the value of $\phi$ for the initial conditions
\begin{displaymath}
  \phi = P(a_0) = 5.315 - 0.44426 a_0 + 0.029593 a_0^2.
\end{displaymath}
The chaotic saddles depicted in Figs. \ref{fig2}(c) and (d) have ``gaps'' clearly distinguished as empty spaces that widen with increasing $\nu_2$. After IC, these gaps are densely filled by the sudden appearance of a special set of UPOs termed coupling UPOs, because they have components in the surrounding and the band regions of the phase space, coupling them \cite{rempel_etal:2004a, szabo_etal:2000, szabo_etal:1996}. This gap filling is an example of a bifurcation phenomenon called ``explosion'' whereby new recurrent points suddenly appear at a nonzero distance from any pre-existing recurrent points \cite{robert_etal:2000}. The gap filling is directly responsible for the crisis-induced intermittency observed in Fig. \ref{fig3}(d), because trajectories can ``escape'' from the banded region to the surrounding region via the unstable manifold of the newly created coupling orbits. In Fig. \ref{fig9}(a) we show the post-crisis strong chaotic attractor (SCA, black) superposed by the period-3 mediating UPO (crosses) and its stable manifold (dashed line) obtained using the contour of the stable manifold of the SCS, for $\nu_2 = 6.515 > \nu_{\mathrm{IC}}$. In Fig. \ref{fig9}(b) we plot the BCS (black circles) and the SCS (gray circles) which constitute the so called geometrical ``skeleton'' of the post-crisis attractor \cite{szabo_tel:1994}. Fig. \ref{fig9}(c) shows one example of a period-10 coupling UPO (crosses). From this figure, it is clear that the coupling UPO has Poincar\'e points located in both the surrounding region and the band region. This coupling UPO was detected by applying the Newton method to a large set of candidate points, obtained from the Poincar\'e points of the SCA located within the gaps of either BCS or SCS.

\section{Conclusion} \label{section_conclusion}

In this paper we performed numerical simulations of a model of nonlinear modulational processes in a plasma, which describes the onset of Langmuir turbulence. We demonstrated the existence of shrimp-shaped structures in a two-parameter space. To our knowledge, this is the first time that these structures are detected in a model of nonlinear wave-wave interactions, rendering support to the universal nature of this phenomenon. Using two numerical methods, the PIM-triple method and the sprinkler method, we identified chaotic saddles and showed their evolution in the bifurcation diagram. We focused on two types of intermittency, the type-I Pomeau-Manneville intermittency, related to a saddle-node bifurcation, and the crisis-induced intermittency, related to an interior crisis. Both types of intermittency were shown to display power-law behavior as the control parameter departs from the transition point. After crisis, we were able to detect one example of a coupling unstable periodic orbit connecting two chaotic saddles, directly responsible for the intermittency observed in the time series after crisis.

Nonlinear modulational processes can describe the formation of modulated wave envelopes in plasmas and fluids which are often observed as intermittent ``bursts'' in a time series. For example, in astrophysical plasmas, observational data of radio signals from pulsars display strong bursts that are intermittent in time \cite{deneva_etal:2009,chian_kennel:1983}. Theoretical models of modulational processes can also be used to explain the formation of extreme or ``freak'' waves in oceans, which are isolated, anomalously high-amplitude ocean waves, exceeding at least 2.2 times the significant wave height, which is defined as the average of the highest one-third of nearby waves \cite{mori_etal:2002,janssen:2003}, and can appear intermittently in space or time, generally under rough sea conditions \cite{osborne_etal:2000, andonowati_etal:2007}. In the open ocean, freak waves can be produced by nonlinear self-modulation of a slowly-varying wave train \cite{janssen:2003}. Simple nonlinear solutions of the NLS equation can serve as prototypes of freak waves in the open ocean, as shown by several theoretical works \cite{peregrine:1983,voronovich_etal:2008} and recent laboratory experiments \cite{chabchoub_etal:2011}. A system of two coupled NLS equations were derived by \citet{onorato_etal:2006}, to describe the modulational instability in the ocean. Their model can explain the occurrence of freak waves in crossing sea states. \citet{voronovich_etal:2008} derived a NLS-type equation in which the bottom stress is taken into account, and found that the bottom friction can affect the formation of freak waves via a modulational instability. However, the three travelling waves truncation of Eq. (\ref{eq_schrodinger_2}) imposes some constraints on the applicability of this model to interpret the intermittent features observed in plasmas and fluids. The choice of one pump wave and one pair of daughter waves might be justified in a weakly nonlinear regime, in which other sets of waves are strongly damped. In this regime. low-dimensional models of nonlinear wave-wave interactions can be used as an alternative to describe modulational processes in plasmas and fluids. For example, \citet{sanchez-arriaga_etal:2009a} compared numerical simulations of the derivative nonlinear Schr\"odinger equation with simulations of a reduced model consisting of three travelling waves in a plasma, and found that the reduced model is in agreement with the DNLS when the energy in the system is small and right-hand polarized waves are present. \citet{janssen:2003} showed that, when the wave steepening is sufficiently weak, the theory of nonlinear four-wave interactions is in good agreement with direct simulations of an ensemble of ocean waves using Zakharov equations and nonlinear Schr\"odinger equation, and can explain the formation of freak waves. Hence, the model used in this paper may be useful to describe qualitatively the dynamics of nonlinear modulational interactions. In summary, our work offers a theoretical explanation for the generation of intermittent bursts resulting from the interaction of chaotic saddles and coupling unstable periodic orbits, and provides new insights to the understanding of these nonlinear phenomena commonly observed in plasmas and fluids.

\acknowledgements

This work is supported by CNPq and FAPESP. A.C.L.C. is grateful for the award of a Marie Curie International Incoming Fellowship and the hospitality of Paris Observatory. The authors thank the referee for valuable comments.

\clearpage

\clearpage

\clearpage

\clearpage

\clearpage

\clearpage

\clearpage

\clearpage

\end{document}